# Current and Shot Noise Measurements in a Carbon Nanotube-Based Spin Diode

Christopher A. Merchant and Nina Marković

Department of Physics and Astronomy, Johns Hopkins University

Baltimore, MD 21218

Low-temperature measurements of asymmetric carbon nanotube (CNT) quantum dots are reported. The CNTs are end-contacted with one ferromagnetic and one normal-metal electrode. The measurements show a spin-dependent rectification of the current caused by the asymmetry of the device. This rectification occurs for gate voltages for which the normal-metal lead is resonant with a level of the quantum dot. At the gate voltages at which the current is at the maximum current, a significant decrease in the current shot noise is observed.

Carbon nanotubes[1] are ballistic conductors that have long mean-free paths and spin diffusion lengths[2,3]. CNTs can therefore be used as spacers for spin valves[4-6] or in other systems that require spin coherence over their lengths[7]. Additionally, short CNT sections behave like quantum dots at low temperatures[8]. Short CNT sections are therefore promising materials for charge detection[9], the separation of spin-entangled pairs of electrons[10,11], or the rectification of current based on the electron spin[12,13]. Recently spin-based current rectification, or a spin diode effect[13–18], has been observed in a lateral ferromagnet-CNT-normal metal device[19]. Shot noise measurements can provide an additional window into the dynamics of a system[20]. Noise measurements are useful for examining the correlation of tunneling events in a system, and have been studied for double-barrier tunnel junctions[21-23]. Here we extend upon previous results by considering the current noise of a CNT-based spin diode.

A schematic of our device is given in Figure 1 (a). It consists of a single-walled carbon nanotube (SWNT) grown by chemical vapor deposition[24] and end-contacted to alternating niobium and cobalt electrodes. Devices are measured by applying a bias voltage and a capacatively-coupled gate voltage and measuring the two-terminal current. A representative AFM image of a device is shown in Figure 1 (b). We have also measured the resistance of sections of the niobium electrode as a function of temperature, which is shown in Figure 1 (c). We observe that there the resistance drops to zero around 6.5 K as the niobium electrode becomes superconducting.

Electronic transport measurements were performed at temperatures ranging from 4.2 K to 10K. Figures 2 (a) and (b) show conductance measurements at both the high and low temperature extremes respectively. We observe typical coulomb blockade behavior at both temperatures[25]. From the conductance maps we are able to determine the ratio of capacitances, $\alpha$, for the single electron transistor (SET)[26]. We measure $\alpha = C_G/C_{tot} \sim 0.02$ for this sample, which is similar to other samples we have measured. We also find that at temperatures below the measured value of $T_c$, as in Fig. 2 (b), an additional energy gap of ~ 1meV opens up, stretching the low-conductance region in the bias direction. The energy value of this extra gap corresponds to 2Δ based on our measurements of $T_c$ of the niobium electrode, where Δ is the superconducting energy gap[27]. This compares to a previous measured gap of 2Δ in quantum dots connected to one superconducting and normal-metal lead[28] and 4Δ for two superconducting leads[29].

Figures 3 (a) and (b) give the conductance as a function of bias voltage and gate voltage at 10 K and 4.2 K respectively. We have focused on a range of the gate voltage where the total coulomb blockade values (U + ΔE) are all about the same. This indicates that the spin-degeneracy of the energy levels of the quantum dot has been broken[30]. The additional superconducting energy gap mentioned previously is also visible here for the measurements done at 4.2 K.

In Figures 3 (c) and (d) we plot the conductance as a function of gate voltage at 10K and 4.2 K, respectively. The measurements at 4.2 K are made at a slightly higher bias to compensate for the superconducting gap. Positive-bias data have been scaled by a factor of 1.2 to compensate for the natural asymmetries in the conductance due to either bias offset or asymmetric electrode couplings. The difference between the positive and negative bias conductance is observed to peak for gate voltages where the normal-metal lead is resonant with an energy level on the dot. The data have been fit using a standard lineshape model[31] with the single-junction resonance locations in the gate voltage and the maximum tunneling rates of the junctions as the fitting parameters.

The single-junction conductances used to fit the overall conductance trace at 4.2 K are given in Figure 4 (a). Positive and negative bias conductance is plotted for both the ferromagnetic and normal-metal tunnel junction. We observe that the normal-metal single junction conductance is symmetric with respect to bias direction, while the ferromagnetic single-junction is not. It has been proposed[13] that this asymmetry is the result of spin accumulation on the dot. The decrease in the tunneling probability through the ferromagnetic tunnel junction leads to the spin diode effect – the current becomes spin polarized for one direction of the bias. We observe the spin diode effect at both temperatures, provided we compensate for the superconducting gap.

In Figure 4 (b) we plot the percent difference between the positive and negative conductance as a function of the bias voltage. It is seen that the difference decreases towards zero as the bias voltage decreases. This means that the spin diode effect is suppressed for low bias voltages as the coulomb blockade essentially overwhelms all other interactions.

A representative plot of the current noise, S, as a function of frequency for several bias voltages is given in Figure 5 (a) for a fixed gate voltage. These noise measurements were made by sampling the fluctuation of the current at 100 kHz using two low-noise amplifiers. The data from the two amplifiers were cross-correlated to reduce amplifier noise[32]. The final noise power was obtained by dividing the power spectral density of the current fluctuations by the white noise response of the system[33]. For low frequencies we see that 1/f noise dominates[34], while for frequencies above ~ 20 kHz we observe the frequency-independent noise, $S_I$, which is characteristic of the shot noise[35].

The shot noise, $S_I$, is plotted against the current in Figure 5 (b) for one gate voltage. We observe that the noise is proportional to the mean value of the current over a range of the bias voltage. We compute the Fano factor[36], F, such that $F = S_I/2eI$, where e is the electron charge and I is the mean current for a given gate and bias voltage. Even though the noise is seen to be asymmetric in Fig. 5 (b), this asymmetry is primarily due to the nature of the quantum dot. It is well known that the noise is determined by the tunneling rates of the individual junctions[37], and so depends strongly on the gate voltage.

A plot of $S_I$ as a function of the gate voltage for ±5 mV bias is given in Figure 5 (c). The theoretical value of the Fano factor has been computed[37] according to the conductance data using F =

$(\Gamma_L^2 + \Gamma_R^2) / (\Gamma_L + \Gamma_R)^2$. For positive bias, we see that the theoretical value for F matches the theory. For negative bias, we observe that the noise dips well below the minimum value of F = 0.5 for certain values of the gate voltage.

In Figure 6 we plot the noise dips below the minimum value of F = 0.5 as a function of the gate voltage. The average value for the minima of the noise dips is measured to be F = 0.28 ± 0.07. Also plotted (right-hand axis) is the current as a function of the gate voltage. We find that the minima of F occur at the gate voltages for which the current reaches a maximum. Values of F below 0.5 are not expected under the standard model for the shot noise[38]. It has been suggested that due to spin flip events on the dot, the Fano factor could be reduced below the F = 0.5 limit[13,39]. However, including spin flip events still does not seem to yield low enough values of F to explain our result. The predicted values are based on the polarization of the cobalt electrode and the absence of polarization in the normal electrode. It is possible that there is an 'effective polarization' which is different than expected due to the spin flip processes. Further study on different systems would be needed to resolve these questions.

It should also be pointed out that super-Poissonian noise might be expected as a result of both spin[13,18,39] and charge[40] accumulation. We find no evidence of this from Fig. 5 (c). The predictions of super-Poissonian noise, however, depend largely on bias and gate voltages, as well as on the spin relaxation times, and it is possible that our samples are not in the right regime to observe this.

In conclusion, we have demonstrated a carbon nanotube-based spin diode. With appropriate tuning of the gate and bias voltages, we are able to turn the spin diode effect on or off. We observe that the effect is highest when a single energy level on the dot is near the Fermi level of the normal electrode. The effect is reduced at low bias where the coulomb blockade effect dominates. We have also seen a significant periodic reduction in the current shot noise, which may be due to increased spin-flip rates at certain gate voltages.

We thank J.C. Egues for useful comments. This work was supported in part by the National Science Foundation under grants No. ECCS-0403964, DMR-0547834 (CAREER) and DMR-0520491 (MRSEC), Alfred P. Sloan Foundation under grant BR-4380, and ACS PRF # 42952-G10.

FIG. 1. (a) Schematic of the measurement setup. Bias is applied between the niobium and the cobalt lead with niobium lead grounded. Gate voltage is applied to the doped silicon substrate through a 500nm thick thermally grown layer of $SiO_2$. DC current and current fluctuations are converted to voltages by a low-noise amplifier and are recorded by a computer-controlled data acquisition (DAQ) board. (b) Atomic force microscope image of a representative sample (scale bar: 1μm). The distance between the electrodes is 300nm. CNT height is measured as < 2 nm. The electrodes are cobalt and niobium, as indicated on the image. (c) Resistance as a function of temperature of a section of the niobium electrode. Transition temperature occurs at 6.5 K.

FIG. 2. Differential conductance, $G=dI/dV$, for varying gate and bias voltages. Scale bar gives conductance values from a minimum of zero (black) up to a maximum of 0.01 $e^2/h$ (blue). Temperature during measurement is 10 K (a) and 4.2 K (b).

FIG. 3. (a) and (b) Differential conductance, $G=dI/dV$, for varying gate and bias voltages. Conductance values range from zero (black) up to 0.06 $e^2/h$ (white). Temperature during measurement is 10 K (a) and 4.2 K (b). (c) Conductance as a function of gate voltage at ±5 mV and 10 K for negative (open red circle) and positive (blue x) bias. Positive bias data have been scaled by a factor of 1.2 and shifted in the gate voltage (~ 0.2 V) to line up conductance peaks. (d) Conductance as a function of gate voltage at ±5.5 mV and 4.2 K for negative (open red circle) and positive (blue x). Positive bias data have been scaled and shifted by the same amounts as (c).

FIG. 4. (a) Single-junction conductance as a function of gate voltage. Nomal-metal conductance is approximately the same for positive (dashed green) and negative (solid green) bias. Ferromagnetic

single-junction conductance is shifted by 0.05 $e^2/h$ for readability. Positive bias (dashed black) conductance is lower than negative bias (solid black) at the gate voltages Vg = 8.85 and 9.25 V. (b) Percent difference between positive and negative bias ferromagnetic single-junction conductance as a function of bias voltage. Data (red circle) is taken at a fixed gate voltage of 8.85 V. A guide for the eye is given as a dashed black line.

FIG. 5: (a) Plot of current noise vs. frequency for varying values of bias voltage. Dashed lines are fits to white noise component above ~ 20 kHz. (b) Representative plot of shot noise vs. current for a fixed gate voltage. Data (open red squares) are plotted alongside theoretical Fano factor values (dashed black lines), F = 0.5 and F = 1.0. The size of the squares represents the error bars of the measured Fano factor values. (c) Fano factor data vs. gate voltage (red open squares) and the theoretical fit based on the conductance data (red line) for positive bias. Positive bias values have been shifted in the gate voltage (~ 0.2 V) to match resonance peaks. (d) Fano factor data vs. gate voltage (blue open squares) and the theoretical fit based on the conductance data (blue line) for negative bias.

FIG. 6: Fano factor dips below F = 0.5 as a function of gate voltage. Fano factor values (black) are labeled on the right axis. Current (dashed red) is plotted as a function of gate voltage and labeled on the left axis. Fano factor dips and current maxima occur at the same gate voltages Vg = 8.75 and 9.15 V. The black squares represent the size of the error bars.

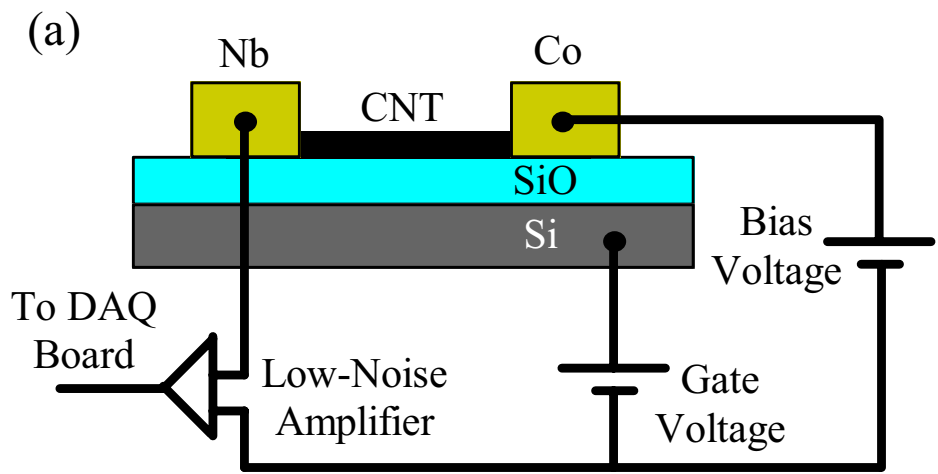

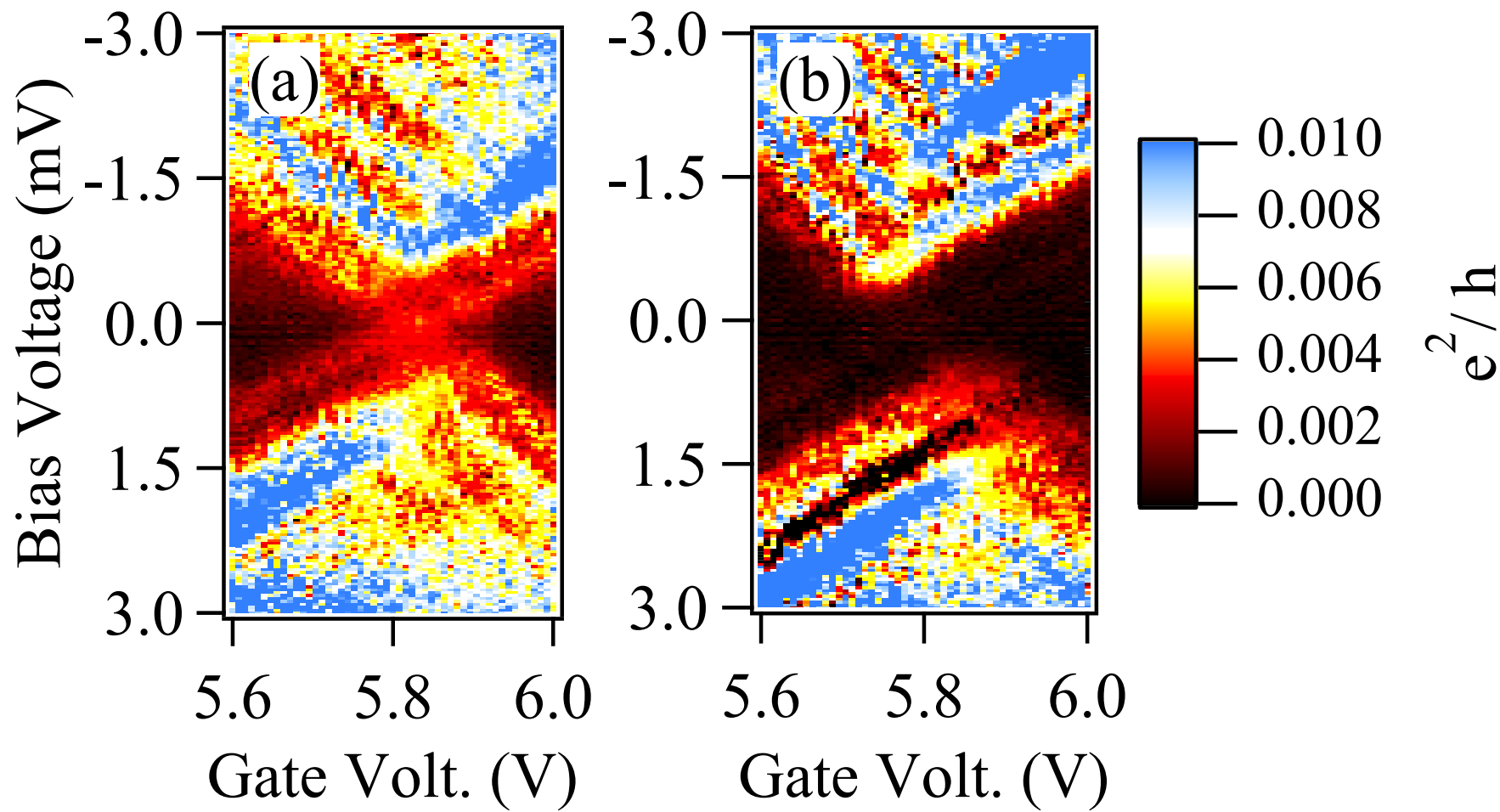

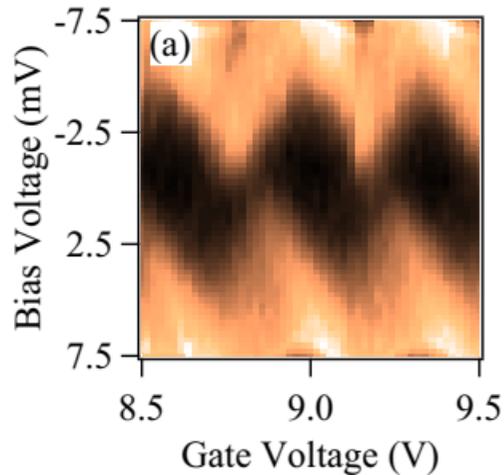
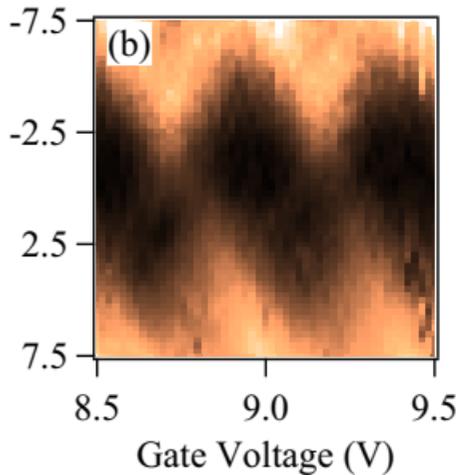
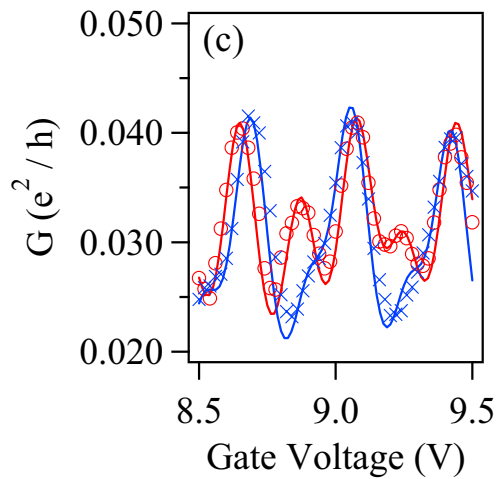
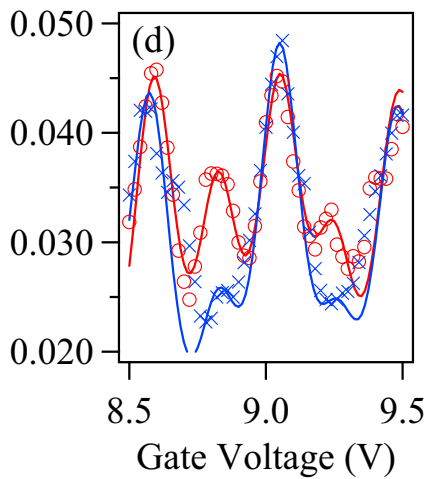

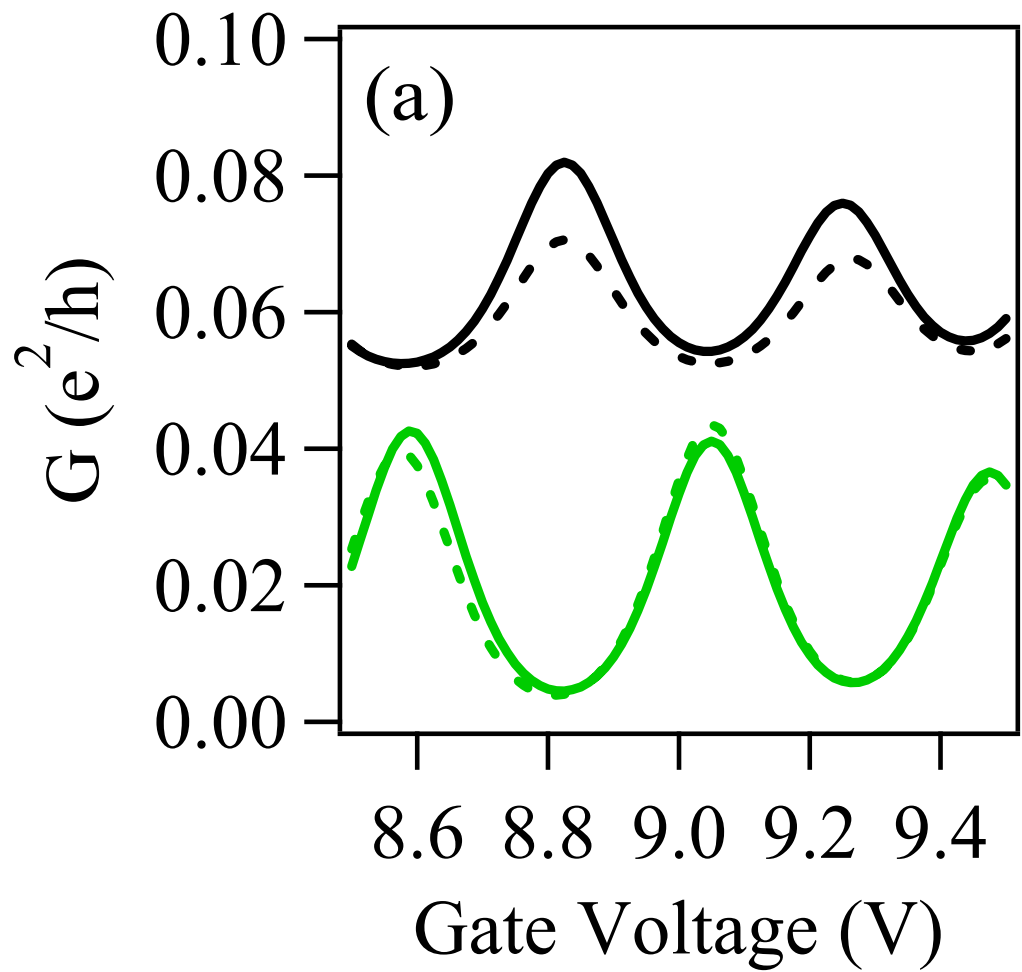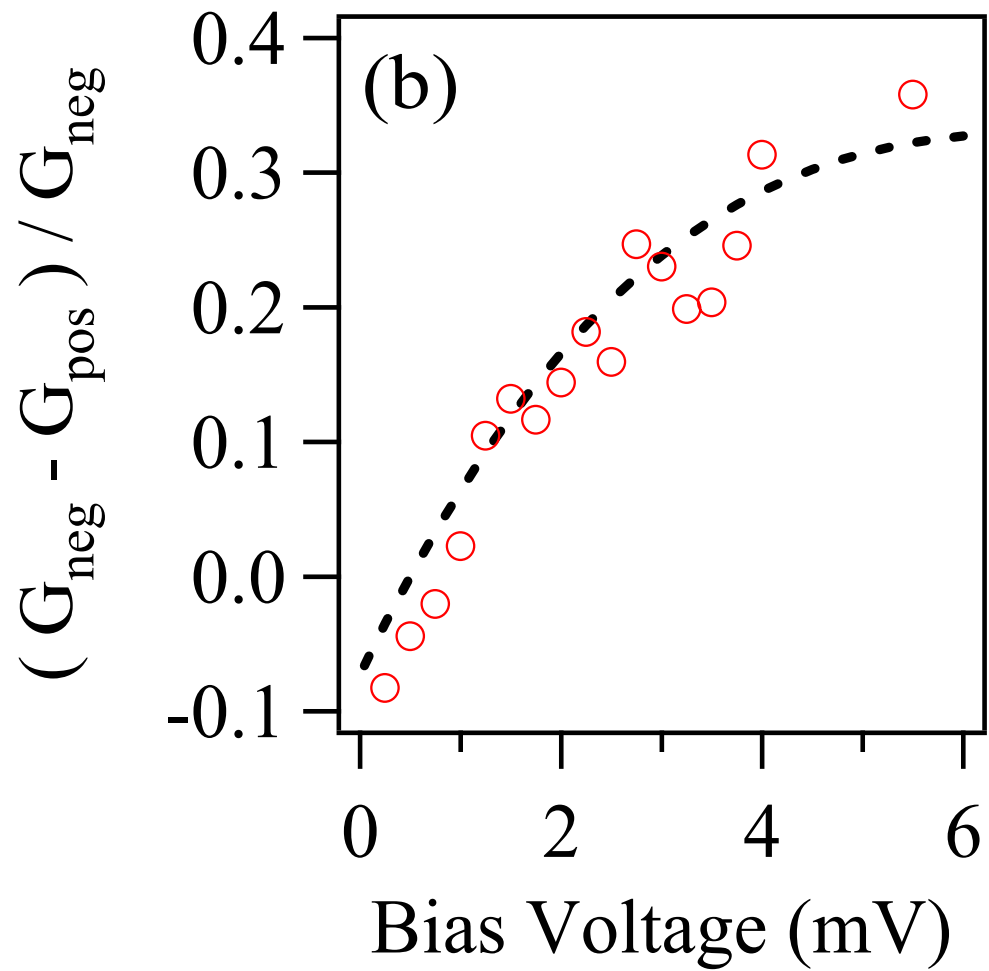

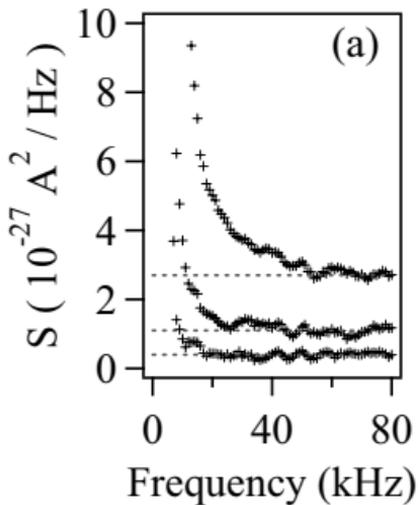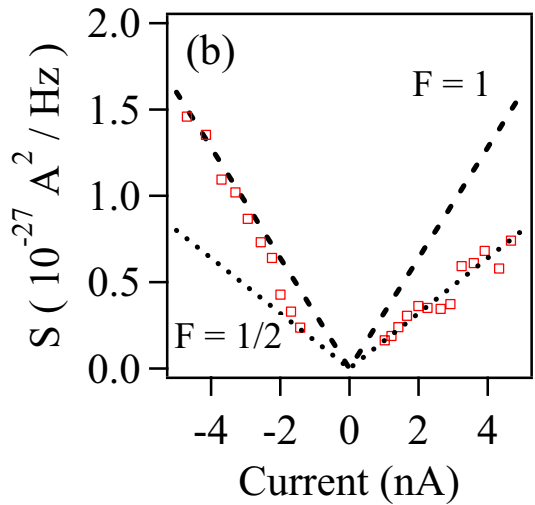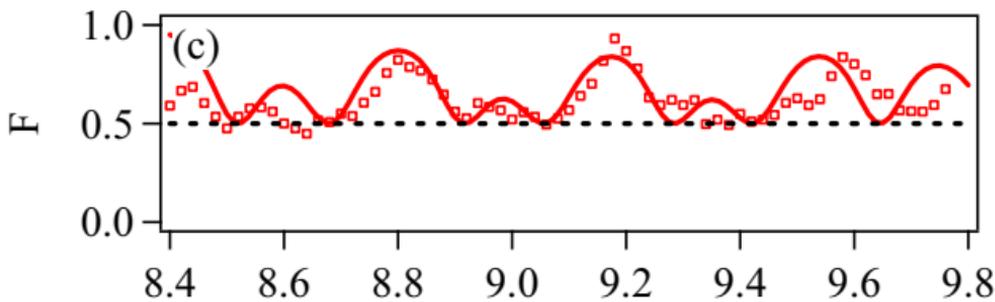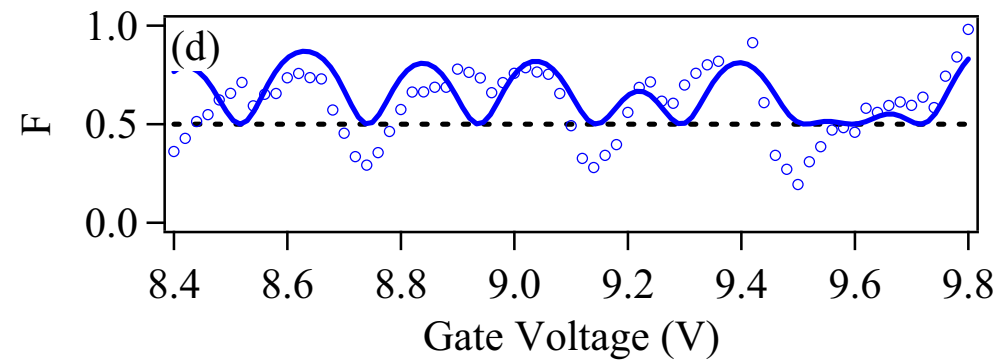

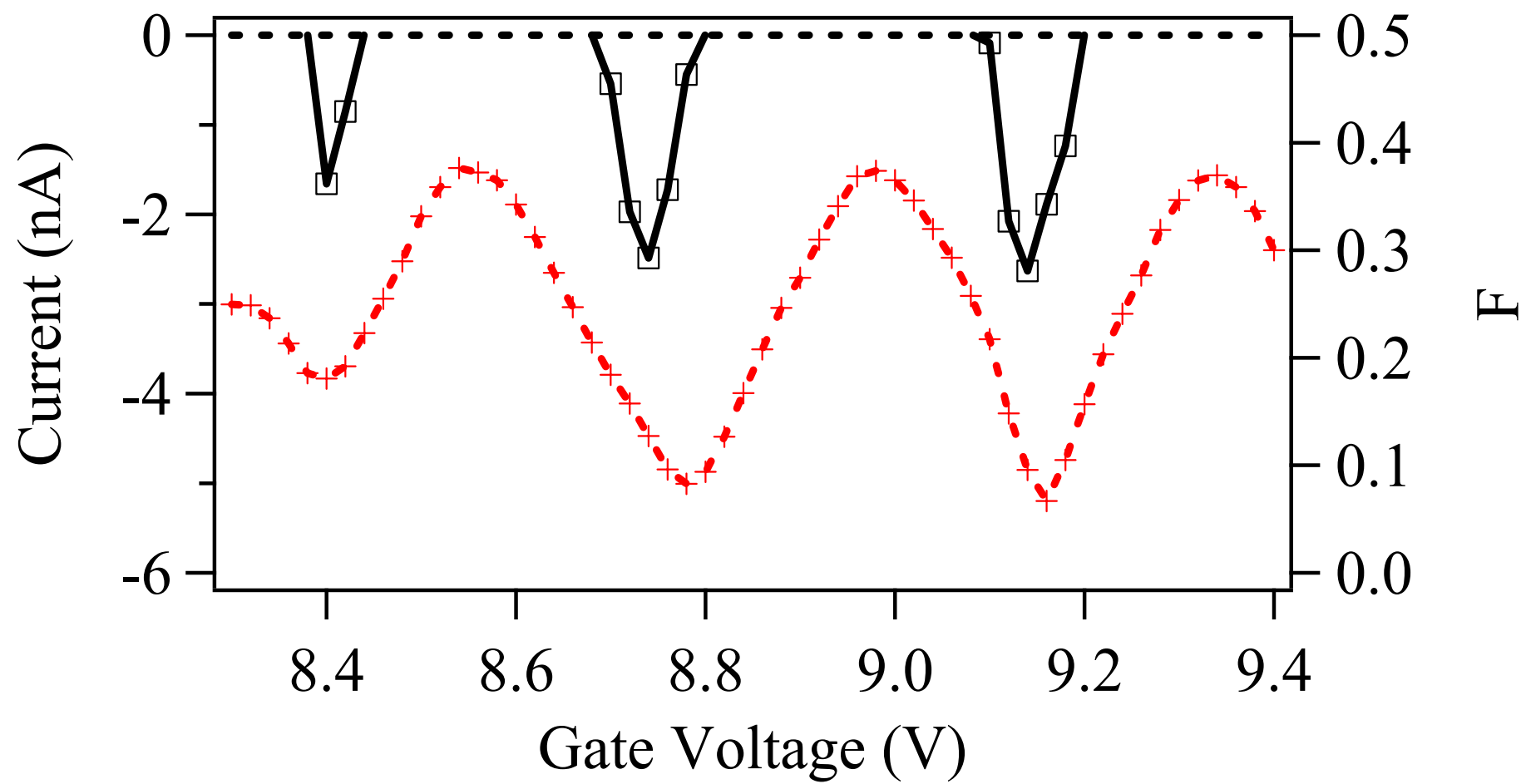